\documentclass[%
 reprint,%
]{revtex4-1}
\usepackage{graphicx}
\usepackage{dcolumn}
\usepackage{bm}
\usepackage[utf8]{inputenc}
\usepackage[T1]{fontenc}
\usepackage{mathptmx}
\usepackage{enumerate}
\usepackage{booktabs}
\usepackage{tabularx}
\usepackage{multirow}
\usepackage[mathscr]{eucal}
\usepackage{amsmath,amssymb}
\usepackage{mathtools}
\usepackage{subfigure}
\usepackage[english]{babel}
\usepackage{lmodern}
\newtheorem{theorem}{Theorem}

\newtheorem{proposition}{Proposition}

\newcommand\mvector{\boldsymbol}
\newcommand\vd{\mvector{d}}
\newcommand\vp{\mvector{p}}
\newcommand\vq{\mvector{q}}

\newcommand\vD{\mvector{D}}

\newcommand\vP{\mvector{P}}

\newcommand\vxi{\mvector{\xi}}
\newcommand\vx{\mvector{x}}
\newcommand\vy{\mvector{y}}

\newcommand\veta{\mvector{\eta}}

\newcommand\scH{{\mathscr H}}
\newcommand\scJ{{\mathscr J}}
\newcommand\scI{{\mathscr I}}
\newcommand\field{\mathbb}

\newcommand\R{\field{R}}

\newcommand\C{\field{C}}
\newcommand\Z{\field{Z}}

\newcommand\vzero{\mvector{0}}
\newcommand\cH{{\mathcal H}}
\newcommand\cF{{\mathcal F}}
\newcommand\rmd{\mathrm{d}}

\newcommand\rmi{\mathrm{i}}

\newcommand\mtext[1]{\quad\text{#1}\quad}
\newcommand\abs[1]{\lvert #1 \rvert}

\newcommand\defset[2]{\left\{{#1}\;\vert \;\; {#2} \,\right\}}
\newcommand\Dt{\frac{\mathrm{d}\phantom{t} }{\mathrm{d}\mspace{1mu}
t}}
\newcommand\tr{\operatorname{Tr}}
\begin{document}
\mathtoolsset{
mathic 
}
\title{Destructive relativity}
\author{Maria Przybylska}
\email{m.przybylska@if.uz.zgora.pl}
\author{Wojciech Szumi{\'n}ski }
\email{w.szuminski@if.uz.zgora.pl}
\affiliation{Institute of Physics, University of Zielona G\'ora, \\
 Licealna 9, 65--417,  Zielona G\'ora, Poland}
\author{Andrzej J. Maciejewski}
\email{a.maciejewski@ia.uz.zgora.pl}
\affiliation{Janusz Gil Institute of Astronomy, University of Zielona G\'ora, \\
Licealna~9, 65-417,  Zielona G\'ora, Poland}

\begin{abstract}
The description of dynamics for high-energy particles  requires an application  of the special relativity theory framework, and  analysis of properties of the corresponding equations of motion is very important.  Here, we analyse Hamilton equations
of motion in the limit of weak external field when potential satisfies the condition $2V(\vq)\ll mc^2$. We formulate very strong necessary integrability conditions for the case when the potential is a homogeneous function of coordinates of integer non-zero
degree. If Hamilton equations are integrable in the Liouville sense, then eigenvalues of the scaled Hessian matrix $\gamma^{-1}V''(\vd)$ at any non-zero solution $\vd$ of the algebraic system $V'(\vd)=\gamma\vd$ must be integer numbers of appropriate form depending on $k$. As it turns out, these conditions are much stronger than those for the corresponding non-relativistic Hamilton equations. According to our best knowledge, the obtained results are the first  general integrability necessary conditions for relativistic systems. Moreover,  a relation between the integrability of these systems and corresponding non-relativistic systems is discussed. The obtained integrability conditions are very easy to use because the calculations reduce to linear algebra.  We show their strength on the example of Hamiltonian systems with two degrees of freedom with polynomial homogeneous potentials. It seems that the only integrable relativistic systems with such potentials are those depending  only on  one coordinate, or  having a radial form. 

The paper has been already published in  ,,Chaos: An Interdisciplinary Journal of Nonlinear Science'', and the final  journal version is available under the link: \href{https://doi.org/10.1063/5.0140633}{https://doi.org/10.1063/5.0140633}.
\end{abstract}

\maketitle
%
\begin{quotation}
Relativistic Hamiltonian equations describing a motion of a point mass   in an arbitrary homogeneous potential are considered.
For the first time,  the necessary integrability conditions for integrability in the Liouville sense for this  class of  systems  are formulated.
These conditions are obtained by means of an analysis of the differential Galois groups of variational equations. They are simple and effective in applications. For instance,  
an application of the necessary integrability conditions for systems with two degrees of freedom shows that relativity  almost completely  destroys integrability, that is, in almost all cases relativistic versions of integrable systems are not integrable.
\end{quotation}

\section{Introduction}
\label{sec:intro}

The study  of the dynamics of classical systems in relativistic regimes is currently in
a great activity of the scientific community. Let us recall the relativistic
Kapitza system~\cite{Guha:21::}; the relativistic hydrogen-like atom in a
magnetic field~\cite{Friedrich:89::, Avazbaev:06::}; the relativistic
two-dimensional harmonic and anharmonic oscillators in a uniform gravitational
field~\cite{Babusci:13::, Vieira:18::,Tung:21::}; the relativistic Lienard-type
oscillators~\cite{Akta:20::} or the relativistic time-dependent
Ermakov--Milne--Pinney systems~\cite{Haas:21::} to mention by name just a few. For more examples see
also~\cite{Bernal:22::,Bernal:18::,Nieto:18::,Fernandez:20::}.

Relativistic systems are an interesting area of research by their own nature
and for their verified applications in many experimental contexts. For instance,
let us mention the recent experimental realization of the harmonic oscillator in
the relativistic regime, using the Bose-condensed lithium atoms in a
two-dimensional optical lattice~\cite{Fujiwara_2018}.

In the context of the classical and non-relativistic systems chaos occurs due to their
inherent nonlinearity of force fields and interactions of systems with these fields. Therefore, in the
simplest case models of a particle moving in flat spaces in a force field with
quadratic or separable potential are integrable. However, the addition of relativistic corrections to an integrable system can destroy its integrability
causing  its  chaotic behavior. In relativistic models the non-integrability
comes from the  modification of the kinetic part of the Hamiltonian -- even if
there are no non-linearities in the potential. Numerical studies, presented
in~\cite{Vieira:18::}, show that the classical two-dimensional Duffing-like
oscillators given by separable potentials of degree four are indeed chaotic.
Therefore, it seems that integrability in special relativity is not only related to the form of the potential.  Hence, there is a natural question which integrable, Newtonian potentials have their integrable counterparts in special relativity and how to distinguish them.

The relativistic versions of classical systems can be considered as
perturbations of the latter. Thus, it is expected that many properties of
classical systems can be destroyed by relativity. Integrability is  such a fundamental property. Hence, one can suspect that relativistic versions of
integrable models are not integrable. The question is what remains. In other words, which systems are integrable in  classical and in  relativistic settings?

The aim of the paper is a study the dynamics and integrability of a
relativistic particle moving in an external potential $V(\vq)$ in the limit of
weak external field, i.e., when $2V(\vq)\ll mc^2$. The considered relativistic
Hamiltonian takes the form
\begin{equation}
  H=mc^2\sqrt{1+\frac{\abs{\vp}^2}{m^2c^2}}+V(\vq),
  \label{eq:ham}
\end{equation}
see e.g. \cite[Sec. 8.4]{Goldstein:02::}, where $\vq=(q_1,\ldots, q_n)\in\R^n,\vp=(p_1, \ldots, p_n)\in\R^n$,  while $m$ is the particle
rest mass and $c$ is the speed of light. For further consideration, we fix
units in such a way that $m=1$ and $c=1$. Moreover, we do not restrict the dimension
of the configuration space, so $n$ is an arbitrary positive integer. The canonical Hamilton's equations
are as follows
\begin{equation}
  \label{eq:eqham}
  \Dt \vq=\frac{\vp}{\sqrt{1+\abs{\vp}^2}}, \qquad
  \Dt \vp =-V'(\vq),
\end{equation}
where $V'(\vq)$ denotes the gradient of $V(\vq)$. Detailed derivation of these equations one can find in \cite{Chanda:18::}, and  for applications, see for instance~\cite{Bernal:18::,Bernal:22::,Vieira:18::,Gomes:22::}.

The plan of the paper is the following. In Section~\ref{sec:intro} the
motivation and main aims of studies of the considered Hamiltonian systems are
given. Section~\ref{sec:numanal} shows results of numerical analysis of famous dynamical
systems: the Kepler problem,  the isotropic and anisotropic oscillators, and the H\'enon-Heiles system in
relativistic and non-relativistic versions. In Section~\ref{sec:main} main
integrability results are presented: about the relation between the integrability of
relativistic and the corresponding non-relativistic systems formulated in
Proposition~\ref{prop:relatnonrelat}, about differential Galois
integrability obstructions given in Theorem~\ref{thm:mo} and results of applications of
these obstructions to non-relativistic Hamiltonian systems with homogeneous
potentials presented in Theorem~\ref{thm:moraleshomo}. Our main theorem devoted to
the integrability of relativistic Hamiltonian systems with homogeneous
potentials is formulated in Theorem~\ref{thm:main} at the end of this section.
Section~\ref{sec:outline} is devoted to the outline of the proof of 
Theorem~\ref{thm:main}. Very strong necessary integrability conditions were
possible to formulate thanks to three types of obstructions: these obtained from an analysis of variational equations on two non-equivalent energy levels studied in
Subsections~\ref{ssec:genericlevel} and \ref{ssec:special} and those for
non-relativistic homogeneous potentials, that  joined together in
Subsection~\ref{sec:final} complete the proof. Section~\ref{sec:applications}
shows the application of the obtained integrability conditions to relativistic
Hamiltonian systems with two degrees of freedom with homogeneous polynomial
potentials. It also explains numerical results presented in Section~\ref{sec:numanal} that
except for radial potentials the passage from non-relativistic to relativistic versions
destroys integrability and for radial potentials, its super-integrability also
seems to be lost.  Final remarks and comments are given in
Section~\ref{sec:recom}. Appendix~\ref{app:riemann} contains basic information
about Riemann~$P$-equation used in Subsection~\ref{ssec:special}.

\section{Numerical analysis of certain potentials}
\label{sec:numanal}

\begin{figure*}[htp]
  \begin{minipage}[c]{0.73\textwidth}
   \subfigure[The classical case]{
      \includegraphics[width=0.45\textwidth]{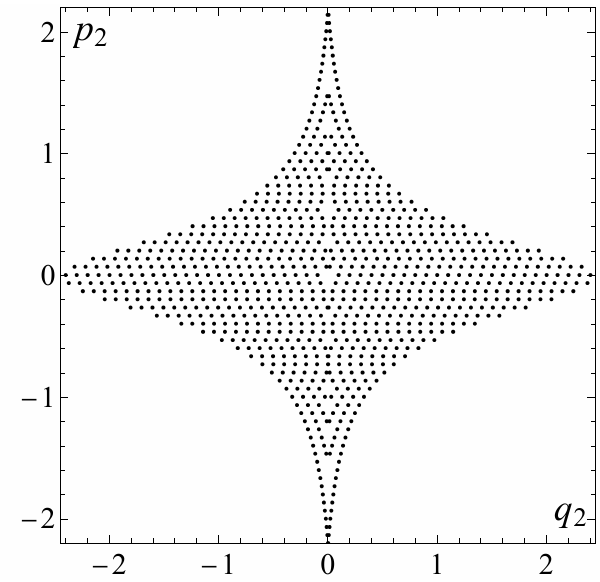}}\hspace{-.1cm}
    \subfigure[The relativistic case]{
  \includegraphics[width=0.45\textwidth]{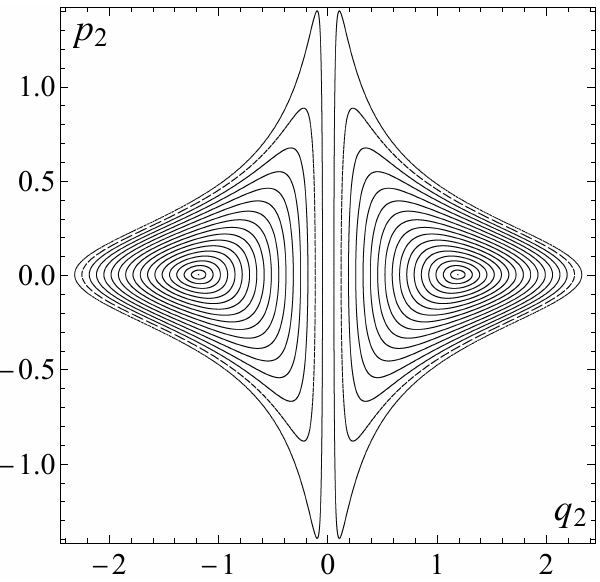}}
  \end{minipage}\hfill
  \begin{minipage}[c]{0.25\textwidth}
\caption{Poincar\'e sections of the classical and the relativistic Kepler
problem made for $\mu=-1/4$ at the energy levels $E=-0.1$ and $E=0.9$,
respectively. The cross-section plane was specified as $q_1=0$, with $ p_1>0$. The
plots present regular dynamics of the systems.  \label{fig:poin_kepler_rel}}
  \end{minipage}
\end{figure*}
In this section, we perform a numerical analysis of classical Hamiltonian
systems and their corresponding relativistic versions with the help of
Poincar\'e cross-sections method. We  show that  relativistic and non-relativistic versions of the isotropic harmonic oscillator and the classical Kepler problem are
integrable and super-integrable, respectively. However, the integrable  anisotropic   harmonic oscillator and the H\'enon-Heiles system  show chaotic trajectories in the relativistic regime.

\subsection{The Kepler problem}
As the first model, we consider the classical Kepler problem, which is governed
by the following Hamiltonian
\begin{equation}
    H=\frac{1}{2}\left(p_1^2+p_2^2\right)+\frac{\mu}{\sqrt{q_1^2+q_2^2}},
\end{equation}
where $\mu\in \R$.  For $\mu<0$ the force is attractive, otherwise, it is repulsive.  Its relativistic counterpart has the form
\begin{equation}
\label{eq:kepler_rel}
    H=\sqrt{1+p_1^2+p_2^2}+\frac{\mu}{\sqrt{q_1^2+q_2^2}}.
\end{equation}
Fig.~\ref{fig:poin_kepler_rel} presents a pair of Poincar\'e sections for the
classical and the relativistic Kepler problem made for $\mu=-1/4$ at the respective energy levels $E=-0.1$ and $E=0.9$. Plots, visible
in~Fig.~\ref{fig:poin_kepler_rel}, show intersections of trajectories
calculated numerically with the suitably chosen surface of section $q_1=0$ and the
direction $p_1>0$. From Fig.~\ref{fig:poin_kepler_rel}(a), one can notice that
all orbits are closed and hence the motion is periodic. Each point corresponds to
a distinct initial condition.  This is due to the  fact that the classical Kepler
problem is maximally super-integrable. Taking into account the relativistic
correction, we observe that the pattern of Fig.~\ref{fig:poin_kepler_rel}(b) is
still very regular but now the motion is  quasi-periodic. Nevertheless, there is no
presence of chaotic behaviour at all. The relativistic
Hamiltonian~\eqref{eq:kepler_rel}  is, in fact, the integrable system and the additional
first integral is the angular momentum $L=q_1p_2 - p_2 q_1$.

\subsection{The harmonic oscillator}

\begin{figure*}[htp]
  \begin{minipage}[c]{0.73\textwidth}
    \subfigure[The classical case]{
      \includegraphics[width=0.45\textwidth]{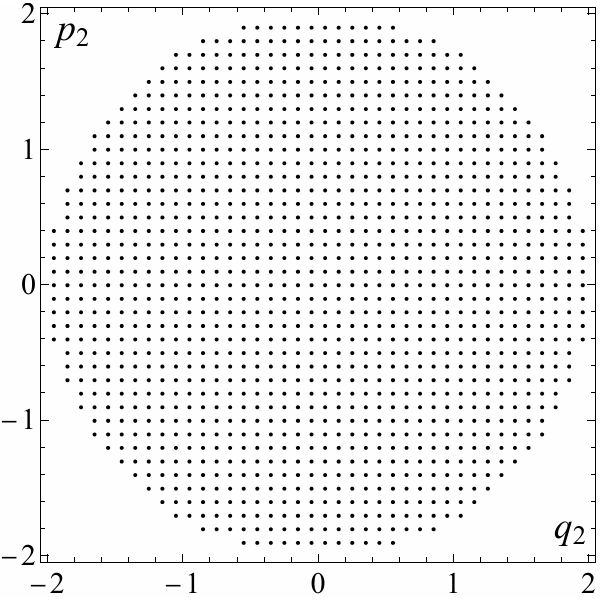}}\hspace{-.1cm}
    \subfigure[The relativistic case]{
      \includegraphics[width=0.45\textwidth]{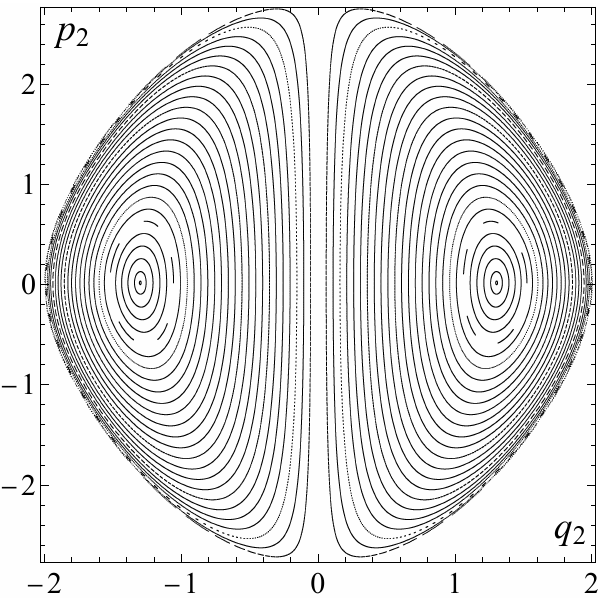}}
  \end{minipage}\hfill
  \begin{minipage}[c]{0.25\textwidth}
    \caption{Poincar\'e sections of the classical and the relativistic isotropic
      oscillator made for $\alpha=1$ at the energy levels $E=E_\mathrm{min}+2$.
      The cross-section plane was specified as $q_1=0$, with $ p_1>0$. The plots
      present regular dynamics of the systems.\label{poin_anizooscillator_rela}}
  \end{minipage}
  \begin{minipage}[c]{0.73\textwidth}
    \subfigure[The classical case]{
      \includegraphics[width=0.45\textwidth]{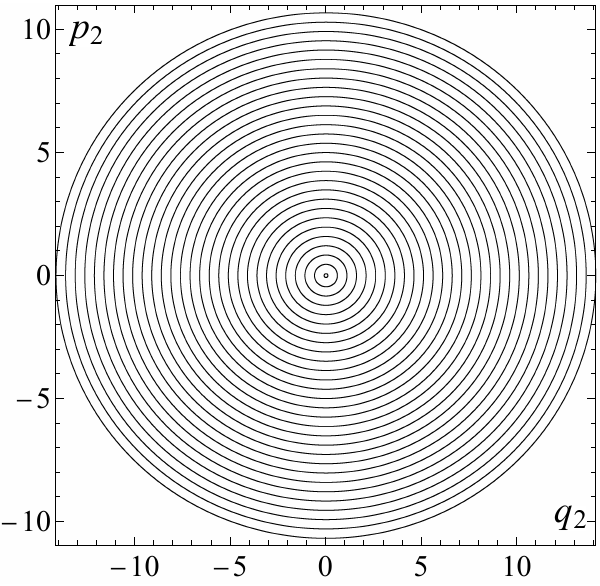}}\hspace{-.1cm}
    \subfigure[The relativistic case]{
      \includegraphics[width=0.45\textwidth]{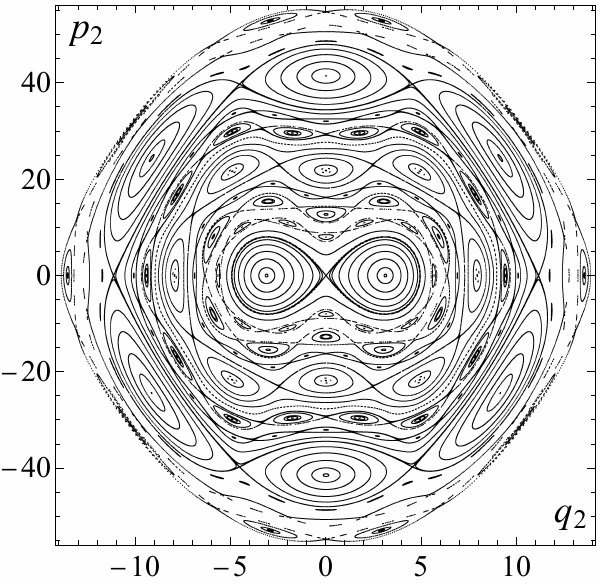}}
  \end{minipage}\hfill
  \begin{minipage}[c]{0.25\textwidth}
    \caption{Poincar\'e sections of the classical and the relativistic
      anisotropic oscillator made for $\alpha=1/2$ at the energy levels
      $E=E_\mathrm{min}+58$. The cross-section plane was specified as $q_1=0$, with
      $ p_1>0$. The plots present regular dynamics of the non-relativistic case
      and chaotic behaviour of the relativistic system.  \label{poin_anizooscillator_relb}}
  \end{minipage}
\end{figure*}

As the second model, we consider 2D harmonic oscillator, which is governed by
the following Hamiltonian function
\begin{equation}
  \label{eq:harmonic_oscillator}
  H=\frac{1}{2}\left(p_1^2+p_2^2\right)+\frac{1}{2}\left(q_1^2+\alpha q_2^2\right),
\end{equation}
where $\alpha$ is a positive parameter. For $\alpha=1$ the system is isotropic
otherwise it is anisotropic. It is obvious that
Hamiltonian~\eqref{eq:harmonic_oscillator} is integrable due to its
separability in the Cartesian coordinates. The corresponding relativistic harmonic oscillator is defined as
follows
\begin{equation}
  \label{eq:harmonic_oscillator_rel}
  H=\sqrt{1+p_1^2+p_2^2}+\frac{1}{2}\left(q_1^2+\alpha q_2^2\right).
\end{equation}
Fig.~\ref{poin_anizooscillator_rela} presents a pair of Poincar\'e
sections of the classical and the relativistic isotropic harmonic oscillators made
for $\alpha=1$ at  constant energy levels $E=E_\mathrm{min}+2$. Here
$E_\mathrm{min}$ denotes the energy  minimum of   the respective systems. For the non-relativistic
system, we have $E_\mathrm{min}=0$, while for the relativistic case
$E_\mathrm{min}=1$. As the non-relativistic case is known to be
super-integrable, the Poincar\'e sections visible in
Fig.~\ref{poin_anizooscillator_rela}(b)  suggest only the integrability of the
relativistic harmonic oscillator. This is due to the presence of quasi-periodic
orbits. Indeed, one can show that the
Hamiltonian system defined by~\eqref{eq:harmonic_oscillator_rel} is  integrable with the
additional first integral $L=q_1p_2 - p_2 q_1$.

\begin{figure}[t]
  \begin{center}
    \includegraphics[width=0.33\textwidth]{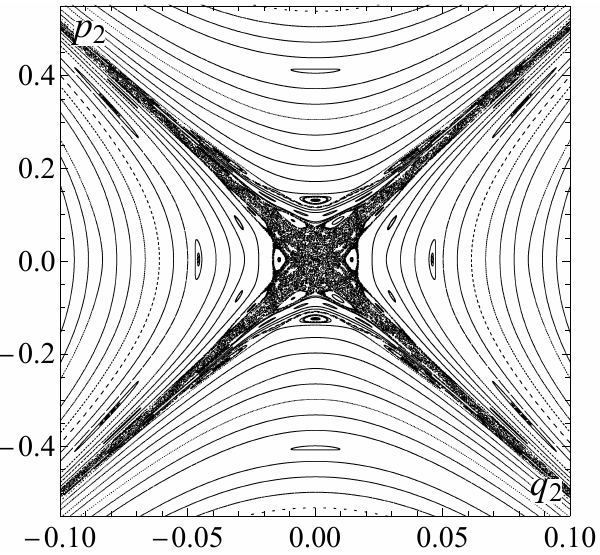}
    \caption{Magnifications of a central part of the Poincar\'e section (see
      Fig.~\ref{poin_anizooscillator_relb}(b)), presenting the chaotic behaviour
      of the system. \label{poin_anizooscillator_relzoom}}
  \end{center}
\end{figure}

Let us consider the anisotropic case. Fig.~\ref{poin_anizooscillator_relb}
presents a pair of Poincar\'e sections for classical and relativistic harmonic
oscillators made for $\alpha=1/2$ at the energy levels $E=E_\mathrm{min}+58$, respectively. In
the non-relativistic case, the Poincar\'e section (see
Fig.~\ref{poin_anizooscillator_relb}(a)) presents integrable motion with one
particular periodic solution surrounded by quasi-periodic orbits. The situation
becomes more complex, when the relativistic correction is taken into
account, see Fig.~\ref{poin_anizooscillator_relb}(b). As we can notice, the
invariant tori become visibly deformed. Some of them are destroyed, and we observe the appearance of stable periodic solutions, which are
enclosed by the separatrices.
Fig.~\ref{poin_anizooscillator_relzoom} shows the magnification of a small part
of the Poincar\'e section in the vicinity of an unstable periodic solution, which
indicates the chaotic nature of the system.

\subsection{The H\'enon--Heiles system}
The classical H\'{e}non-Heiles potential is perhaps one of the most simple,
classical Hamiltonian systems, which can exhibit both integrable and chaotic
dynamics depending on the values of parameters. It is described by means of the following Hamiltonian 
\begin{equation}
  \label{eq:hamiltonian_henon}
  H=\frac{1}{2}\left(p_1^2+p_2^2\right)+\frac{1}{2}\left(q_1^2+q_2^2\right)+\alpha q_1^2q_2+\frac{1}{3}\beta q_2^3,
\end{equation}
where $\alpha,\beta$ are real parameters. This potential appears in various
problems in physics. For instance, in celestial mechanics ~\cite{Henon:64::}, in
statistical and quantum mechanics~\cite{Ford:73::}, and recently it  has been applied also in Hamiltonian neural networks~\cite{Mattheakis:22::}, to cite just
a few.

There are three known integrable cases of the H\'{e}non-Heiles
model~\cite{Chang:82::,Grammaticos:83::},  namely, $\alpha=0$, $\beta/\alpha=1$
and $\beta/\alpha=6$. In all cases, additional first integrals are quadratic
polynomials with respect to the momenta, and the Hamiltonian is separable in
appropriate coordinates. It was proved by Ito in~\cite{Ito:85::} and later
complemented in~\cite{Morales:99::,Li:11::} that the above values of the
parameters are the only ones for which the H\'{e}non-Heiles model is integrable.

\begin{figure*}[htp]
  \begin{minipage}[c]{0.73\textwidth}
   \subfigure[The classical case]{
      \includegraphics[width=0.45\textwidth]{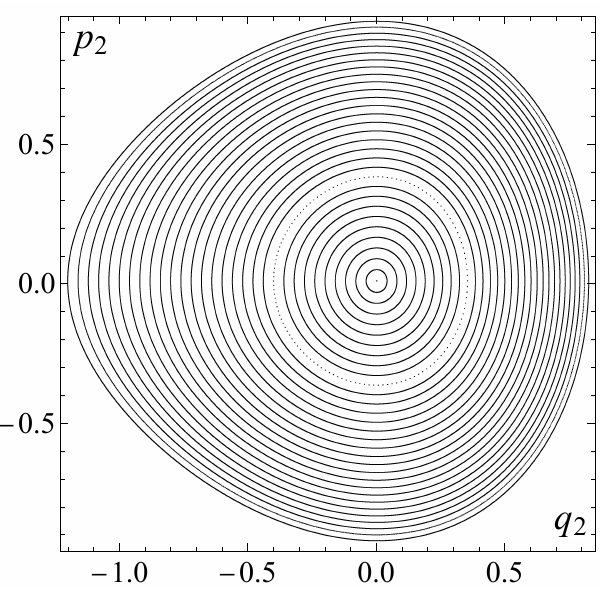}}\hspace{-.1cm}
    \subfigure[The relativistic case]{
  \includegraphics[width=0.45\textwidth]{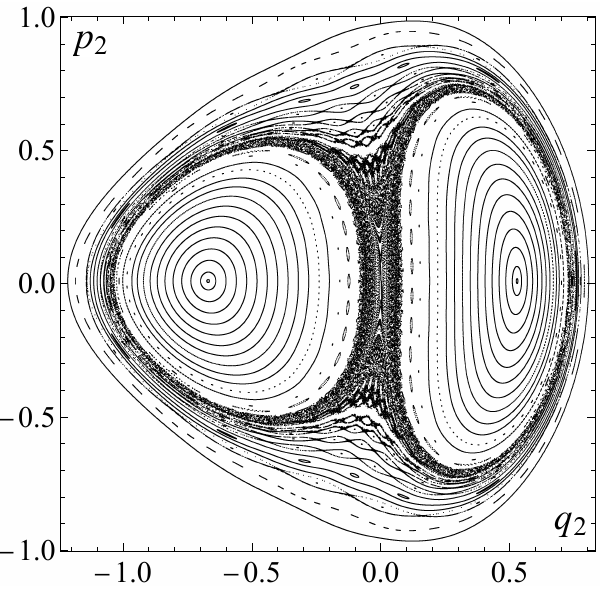}}
  \end{minipage}\hfill
  \begin{minipage}[c]{0.25\textwidth}
      \caption{Poincar\'e sections of the classical and the relativistic
      H\'{e}non-Heiles model made for parameters $\alpha=0$ and $\beta=1/2$ at
      the energy levels $E=E_\mathrm{min}+0.6$. The cross-section plane was specified
      as $q_1=0$, with $ p_1>0$.  The plots
present integrable dynamics of the
non-relativistic case and chaotic
nature of the corresponding relativistic system.\label{poin_henon0}}
       \end{minipage}
  \begin{minipage}[c]{0.73\textwidth}
   \subfigure[The classical case]{
      \includegraphics[width=0.45\textwidth]{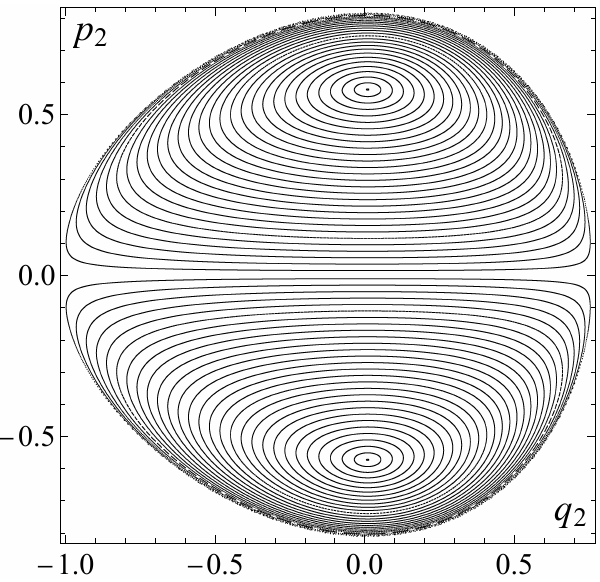}}\hspace{-.1cm}
    \subfigure[The relativistic case]{
  \includegraphics[width=0.45\textwidth]{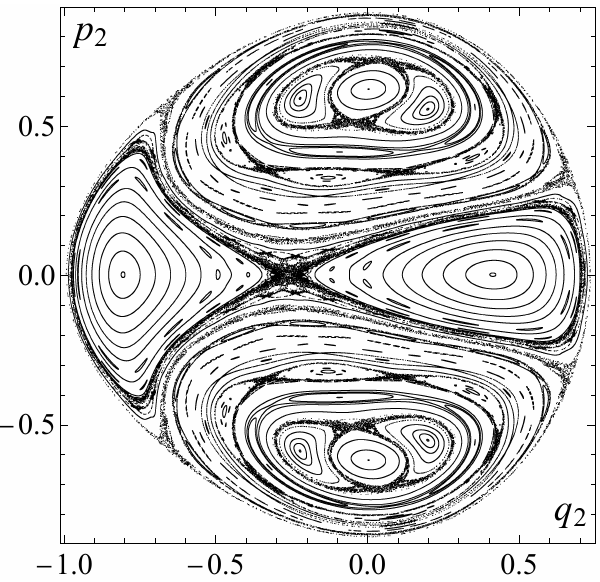}}
  \end{minipage}\hfill
  \begin{minipage}[c]{0.25\textwidth}
      \caption{Poincar\'e sections of the classical and the relativistic
      H\'{e}non-Heiles model made for parameters $\alpha=\beta=1/2$ at the
      energy levels $E=E_\mathrm{min}+0.33$. The cross-section plane was specified as
      $q_1=0$, with $ p_1>0$. The plots
present regular dynamics of the
non-relativistic case and chaotic
dynamics of the relativistic model. \label{poin_henon1}}
       \end{minipage}
        \begin{minipage}[c]{0.73\textwidth}
   \subfigure[The classical case]{
      \includegraphics[width=0.45\textwidth]{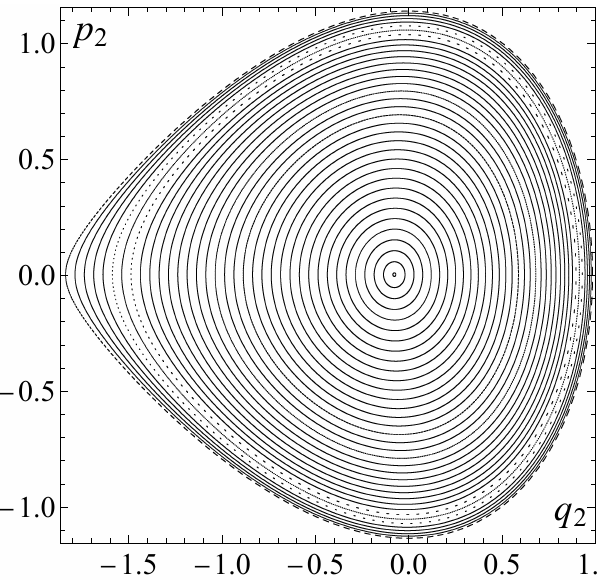}}\hspace{-.1cm}
    \subfigure[The relativistic case]{
  \includegraphics[width=0.45\textwidth]{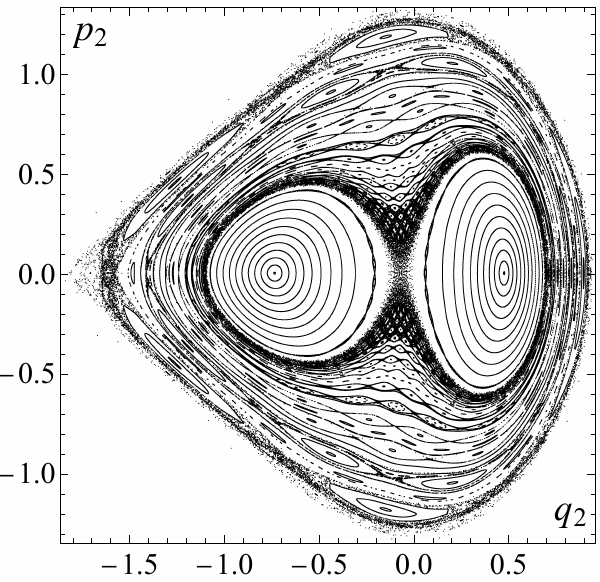}}
  \end{minipage}\hfill
  \begin{minipage}[c]{0.25\textwidth}
    \caption{Poincar\'e sections of the classical and the relativistic
      H\'{e}non-Heiles model made for parameters $\alpha=1/12, \beta=1/2$ at the
      energy levels $E=E_\mathrm{min}+0.7$. The cross-section plane was specified as
      $q_1=0$, with $ p_1>0$.  The plots
show integrable dynamics of the
non-relativistic model and chaotic
nature of the system in the relativistic regime. \label{poin_henon2}}
       \end{minipage}
\end{figure*}

The corresponding relativistic H\'{e}non-Heiles  system  is defined as
follows
\begin{equation}
  \label{eq:Hamiltonian_henon_relativistc}
  H=\sqrt{1+p_1^2+p_2^2}+\frac{1}{2}\left(q_1^2+q_2^2\right)+\alpha q_1^2q_2+\frac{1}{3}\beta q_2^3.
\end{equation}
Figs.~\ref{poin_henon0}-\ref{poin_henon2} show the Poincar\'e sections of the
classical and the relativistic H\'{e}non-Heiles model for values of parameters:
$\alpha=0$, and $\beta/\alpha=1$ and $\beta/\alpha=6$. As we can notice, the
general shapes of the sections' boundaries  are similar, whereas the dynamics presented within
them are completely different. In the non-relativistic cases, we obtain shapely
elegant integrable curves with mostly quasi-periodic solutions. In the relativistic regime,
however, the invariant tori broke up into sequences of stable and unstable
periodic solutions. Moreover, in the neighborhood of unstable periodic
solutions, we observe regions at the planes where chaotic motion takes place.
This suggests the non-integrability of the relativistic H\'{e}non-Heiles model.

\section{Main results}
\label{sec:main}

In the previous section,  with the help of numerical analysis, we have shown that the super-integrable non-relativistic Kepler problem and the harmonic oscillator are integrable in a relativistic regime.  On the other hand, the integrable non-relativistic models of the anisotropic harmonic oscillator and
the H\'enon--Heiles  system seem to be not integrable when the relativistic correction is taken into the account. The above models were just examples and  an effective criterion
for the identification of integrable relativistic systems is needed. This is the aim of the paper.

At first, we make a simple observation. Note that $v=\abs{\dot\vq}=\abs{\vp}/u$,
so the Lorenz factor is $u = 1/\sqrt{1-v^2}$. In the non-relativistic limit
$u\rightarrow 1$. The power series expansion of Hamiltonian~\eqref{eq:ham} gives
\begin{multline}
  \label{eq:ser}
  H=\sqrt{1+\abs{\vp}^2}+V(\vq)=\left(1 +\frac{1}{2}\abs{\vp}^2+
    \cdots\right)+V(\vq)
  \\=1+\cH+\cdots
\end{multline}
where $\cH$ is the non-relativistic natural Hamiltonian
\begin{equation}
  \cH =\frac{1}{2}\abs{\vp}^2+V(\vq).
  \label{eq:hamnon-rel}
\end{equation}
Thus, in the non-relativistic limit the Hamiltonian \eqref{eq:ham}  becomes (after
neglecting the constant term $mc^2=1$) the standard non-relativistic
Hamiltonian.

Investigating a specific system is important to know if it is integrable. In
general, it is difficult to give an answer to this question. Here we formulate this
question for the relativistic version of classical natural systems. It is hard to
expect  a complete answer to this question as it is unknown for
non-relativistic ones. Nevertheless, we propose to study the relativistic versions of
those systems for which integrability was investigated deeply. Here we have in
mind natural systems with homogeneous potentials, which were investigated in
numerous works, see  for instance \cite{Hietarinta:83::,Hietarinta:87::,Ito:85::,Yoshida:87::,Yoshida:89::,Almeida:1998::,Nakagawa:01::,Morales:01::a,Morales:99::c,mp:04::d,mp:05::c,mp:05::d,mp:08::c,mp:09::a,mp:09::b,mp:10::a,mp:10::b,mp:12::a,mp:13::a,mp:15::d,mp:16::a,mp:17::b,Llibre:18::,mp:20::e}.

Let us assume that potential $V(\vq)$ in the relativistic Hamiltonian
\eqref{eq:ham} is a homogeneous function of integer non-zero degree $k$, that is
$V(\lambda \vq)=\lambda^k V(\vq)$ for $\lambda>0$. If we assign weights: $2$ to
coordinates and $k$ to momenta, then in the non-relativistic limit
Hamiltonian~\eqref{eq:hamnon-rel} is weight homogeneous of weight-degree $2k$.
This weight-homogeneity is consistent with the canonical structure. That is, a
non-vanishing Poisson bracket $\{\cH,\cF\}$ with a weight-homogeneous function
of weight-degree $l$, is weight-homogeneous of weight-degree $k+l -2$. Then,
expansion~\eqref{eq:ser} gives a weight-homogeneous expansion of the form
\begin{equation}
  \label{eq:serh}
  H-1 = \sum_{i=0}^\infty H_{2k+i},
\end{equation}
where $H_l$ is a weight-homogeneous function of weight-degree $l$. Thus,
$\scH=H_{2k}$ is the leading term of this expansion. Now, if $H$ admits a certain number of functionally independent first integrals, then the weight-homogeneous leading terms of them are first integrals of $\scH=H_{2k}$. Thanks
to the Ziglin Lemma~\cite{Audin:01::}, we can assume that they are functionally
independent.

From the above, we can deduce the following fact.
\begin{proposition}
  \label{prop:relatnonrelat}
  If the relativistic Hamiltonian system governed by Hamiltonian~\eqref{eq:ham} with
  a homogeneous  potential $V(\vq)$ is integrable in the Liouville sense, then
  its non-relativistic counterpart defined by Hamiltonian $\cH$ given in~\eqref{eq:hamnon-rel} with the same $V(\vq)$ is also integrable in the Liouville sense.

\end{proposition}
It means that the integrability of a non-relativistic Hamiltonian is a necessary
condition  for the integrability of its relativistic counterpart.
Moreover, if potential $V(\vq)$ has expansion
$
V(\vq) = V_k(\vq) + \cdots,
$
where $V_k(\vq)$ is homogeneous of degree $k$, and dots denote terms of homogeneous degrees
higher than $k$, then still we obtain expansion of the form (11). Hence, if the relativistic Hamiltonian
system defined by (1) with a non-homogeneous potential $V(\vq)$ is integrable in the Liouville sense,
then the corresponding non-relativistic system defined by Hamiltonian $\cH$ with potential
$V_k(\vq)$ is also integrable in the Liouville sense.

To prove our main results we apply the Morales--Ramis theory. Its detailed description with many examples one can find in  book~\cite{Morales:99::c}, see also
\cite{Morales:99::b,Morales:01::b1,Morales:01::b2}. The main idea is to
investigate the variational equations of the considered non-linear Hamiltonian system along a particular non-equilibrium solution. Passage to the variational equations, which are linear enables to use the differential Galois group related to them. A first integral of the considered non-linear system  generates  the  first integral of variational equations, which is also an invariant of the Lie algebra of their  differential Galois group. In the case of integrability in the Liouville sense, the number  of first integrals and their commutations implies the Abelianity of the Lie algebra of the differential Galois group and the identity component of this group. This reasoning explains the origin of the fundamental theorem of the Morales--Ramis theory.
\begin{theorem}[Morales,1999]
  \label{thm:mo}
  If a Hamiltonian system is integrable in the Liouville sense in a neighborhood of a particular solution, then the identity component of the
  differential Galois group of the  variational equations along this solution is Abelian.
\end{theorem}
The true strength of the above theorem appears when it is applied to classical
systems~\eqref{eq:hamnon-rel} with homogeneous potentials $V(\vq)$ of integer
degree $k$. For such potentials a non-zero vector $\vd$ satisfying
\begin{equation}
V'(\vd)=\gamma \vd,
\label{eq:darboux}
\end{equation}
for a certain non-zero $\gamma$, is called a Darboux point of this potential.
Let $(\lambda_1, \ldots,\lambda_n)$ denotes the eigenvalues of scaled Hessian
matrix $\gamma^{-1}V''(\vd)$. Vector $\vd$ is its eigenvector. The corresponding
eigenvalue we  denote by $\lambda_n$, and as it is easy to show using homogeneity of the potential that
$\lambda_n=k-1$. Since it does not give obstructions to the integrability it is called  a trivial eigenvalue.
\begin{theorem}
  \label{thm:moraleshomo}
  Assume that the Hamiltonian system defined by Hamiltonian
  \eqref{eq:hamnon-rel} with a homogeneous potential $V(\vq)\in\C(\vq)$ of
  degree $k\in\Z^{\ast}=\Z\setminus\{0\}$ is integrable in the Liouville sense with
  meromorphic first integrals. Then, for each eigenvalue $\lambda$ of
  $\gamma^{-1}V''(\vd)$, pair $(k,\lambda)$ belongs to the following list:
  \begin{equation*}
    \small{\begin{array}{rcr}
        \hline
         k & \lambda  &\\[0.5em]
        \hline
        \pm 2 &  \lambda  & \\[0.9em]
        k & p + \frac{k}{2}p(p-1) & \\[0.9em]
        k & \frac{(k p+1) (k p+k-1)}{2 k}  & \\[0.9em]
        3 & \frac{1}{8} (2 p+1) (6 p+1), & \frac{1}{96} (12 p+1) (12 p+5) \\[0.9em]
        &\frac{1}{600} (30 p+1) (30 p+11),  &
        \frac{1}{600} (30 p+7) (30 p+17) \\[0.9em]
        4 &\frac{1}{72} (12 p+1) (12 p+7) & \\[0.9em]
        5 &\frac{1}{360} (30 p+1) (30 p+19), & \frac{1}{40} (10 p+1) (10 p+7)\\[0.9em]
        -3 &-\frac{1}{8} (2 p-1) (6 p+7), & -\frac{1}{96} (12 p-7) (12 p+13) \\[0.9em]
        & -\frac{1}{600} (30 p-19) (30p+31), & -\frac{1}{600} (30 p-13) (30 p+37)\\[0.9em]
        -4 & -\frac{1}{72} (12 p-5) (12 p+13) & \\[0.9em]
        -5 &-\frac{1}{360} (30 p-11) (30 p+31) ,& -\frac{1}{40} (10 p-3) (10 p+11)
      \end{array}
    }
  \end{equation*}
  where $p$ is an integer and $k\neq 0$.
\end{theorem}
It is really amazing result: testing of the integrability is reduced to purely
algebraic calculations. Notice that, except the case $k=\pm 2$, if the system is
integrable, then all eigenvalues $(\lambda_1, \ldots,\lambda_n)$ are rational
numbers.

For the formulation of our main results, we introduce the following two functions
\begin{multline}
  \label{eq:fpm}
  f_\pm(k,p)= 3 k p (2 p+1)+\\
  \frac{1}{2}\left[1\pm (4 p+1) \sqrt{4 k^2 p (2 p+1)+1}\right],
\end{multline}
and two sets
\begin{equation}
  \label{eq:Jpm}
  \scJ_{\pm}=\defset{f_\pm(k,p)}{p\in\Z}\cap\Z.
\end{equation}
Additionally, let $\mathscr{S}$ denote the set of square triangular numbers.
That is, numbers $s$ such that $s = q^2$ and $s = \tfrac{1}{2}p(p+1)$ for
certain integers $q,p\in\Z$, see e.g. \cite[p. 10]{Dickson:66::}. Then, we denote
\begin{equation}
  \begin{split}
 & \scJ_{1}=\mathscr{S}, \qquad \scJ_{2}= \left\{ \frac{1}{2}p(p+1) \ |\  p\in\Z \right\}, \\
  &\scJ_{-1}=\left\{1-s \ |\ s\in\mathscr{S}\right\}
  , \qquad \scJ_{-2}=\left\{1-s \ |\ s\in \scJ_2\right\}.
  \end{split}
\end{equation}

The main result of our analysis has the following form.
\begin{theorem}
  \label{thm:main}
  Assume that a Hamiltonian system defined by Hamiltonian \eqref{eq:ham}, with
  a homogeneous potential $V(\vq)\in\C(\vq)$ of degree
  $k\in\Z^{\ast}=\Z\setminus\{0\}$, is integrable in the Liouville sense with
  first integrals, which are rational functions of $(\vq,\vp,u)$, where
  $u=\sqrt{1+\vp^2}$.
  Then,
  \begin{itemize}
    \item if $\abs{k}> 2$, each eigenvalue $\lambda=\lambda_i$ of matrix
    $\gamma^{-1}V''(\vd)$ with $i=1,\ldots, n-1$, belongs to the set
    $\scJ_+\cup\scJ_-$;
    \item if $\abs{k}\leq2$, $k\neq0$, each eigenvalue $\lambda$ belongs to the set
    $\scJ_k\cup \scJ_+\cup\scJ_-$.
  \end{itemize}
\end{theorem}
Notice that the necessary condition for the integrability is that all
eigenvalues of $V''(\vd)$ are integers, so it is more restrictive than in the
non-relativistic case. Moreover, and not so evident, these integers are
extremely rare. For example, taking $k=4$, we find that among numbers
$f_\pm(k,p)$ with an integer $\abs{p}\leq 10^6$ only $9$ are integers.

In our Theorem~\ref{thm:main} we require that the first integrals depend on the
additional variable $u$. The reason is the following. We cannot use directly the
Morales--Ramis theory for the study of the integrability of the system governed by
Hamiltonian \eqref{eq:ham} because it was originally formulated for meromorphic Hamiltonians and meromorphic corresponding Hamiltonian fields.  In fact, our Hamiltonian \eqref{eq:ham} is not a meromorphic
function because it contains the term $u=\sqrt{1+\vp^2}$. However, it is an algebraic
function of canonical coordinates. Then, for the extended applicability of the Morales--Ramis
theory for the case of algebraic potentials, we can use an approach proposed in
\cite{Combot:13::,mp:16::a}. Here we just notice that  when we add the additional
variable $u=\sqrt{1+\vp^2}$, then the
relativistic system \eqref{eq:eqham} one can rewrite as a Poisson system with
rational right-hand sides with a polynomial Hamiltonian. Indeed, Hamiltonian~\eqref{eq:ham} takes the form
\[
  \scH=u+V(\vq),
\]
and its corresponding equations of motion~\eqref{eq:eqham} transform into
\begin{equation}
\frac{\mathrm{d}\vq}{\mathrm{d}t}=\frac{1}{ u}\vp, \quad \frac{\mathrm{d}\vp}{\mathrm{d}t} =-V'(\vq),\quad \frac{\mathrm{d}u}{\mathrm{d}t}=-\frac{1}{u}\vp\cdot V'(\vq),
  \label{eq:poisso}
\end{equation}
where we joined the time derivative of the additional variable $u$. This system
can be written as
\[
  \dot \vx=\vP\nabla_{\vx}\scH, \qquad \vx=(\vq,\vp,u),
\]
with the Poisson bivector
\begin{equation}
  \vP=\begin{bmatrix}
    \mathbb{O}&\mathbb{I}&\tfrac{1}{u}\vp\\
    -\mathbb{I}&\mathbb{O}&\vzero\\
    -\tfrac{1}{u}\vp^T&\vzero^T&0
  \end{bmatrix},
\end{equation}
where $\mathbb{O}$ and $\mathbb{I}$ denote the $n\times n$ zero matrix and the
identity matrix, respectively, $\vzero$ is $n$-dimensional column of zeros, $T$
denotes the transpositions and $\nabla_{\vx}\scH=[V'(\vq),\vzero,1]^T$. The
skew-symmetry of this structure is evident, and the Jacobi identity one can check by
direct calculations. This Poisson structure has one Casimir function
$\scI=u^2-\vp^2.$

\section{Outline of the proof}
\label{sec:outline}

Our main result was  formulated as the theorem, so it needs a proof. It is quite
long, so we will not present here it in the whole extent. However, we believe that
for a reader it will be profitable to know  just the basic ideas of this proof.

The starting point in our proof is the Morales-Ramis theorem formulated in Theorem~\ref{thm:mo}. For its effective application, we
need a particular solution of the considered system. In general, there is no
universal method of finding it.  However, for the classical case of natural Hamiltonian
systems with a homogeneous potential, one can find  `straight line' solutions
along a Darboux point of the potential. In the same way, we find particular
solutions for the relativistic version of the system. The variational equations
along such a solution split into independent scalar equations of the second order.

If the system is integrable, then the identity component of the differential
Galois group of each of these scalar equations is Abelian.  To check this
property  we transform each of these equations  into equations with rational
coefficients. This gives us the possibility to use all known results of the classical theory of such
equations \cite{Ince:44::,Poole:60::} as well as the Kovacic algorithm~\cite{Kovacic:86::} and its
numerous improvements~\cite{Duval:92::,Ulmer:96::}.  The difficulty
of the problem is connected to the fact that the variational equations depend
on parameters.

In the relativistic case, the choice of a particular  energy level is
important. By its proper choice, we achieve a confluence of singular points of the considered
equation. Thanks to this, we  obtain  the Riemann $P$-equation for which the
differential Galois group is known. In this way, we get quite strong  necessary
conditions for integrability. For their improvement, we investigated   the
variational equations  for a generic value of the particular solution energy.
The key step in our reasoning is as follows. If  the relativistic model is
integrable, then three types of conditions have to be fulfilled. Namely, the system has
to be integrable in the non-relativistic model, so conditions of
Theorem~\ref{thm:moraleshomo} have to be fulfilled. Moreover, simultaneously,
conditions obtained from the analysis of variational equations for the specific
and generic values of the energy also have to be fulfilled.

\subsection{Particular solution and variational equations}
\label{ssec:partsol}

With an arbitrary  Darboux point $\vd$ of the potential $V(\vq)$, one can associate a straight
line particular solution of the form $\vq(t) = \varphi(t)\vd$, where
$\varphi=\varphi(t)$ is a scalar function. As
$\dot\vq=\dot \varphi(t)\vd$, the corresponding momentum $\vp$ can be calculated
from the first half of the Hamilton equations \eqref{eq:eqham}. Hence, the particular solution
is given by
\begin{equation}
  \label{eq:part}
  \vq(t) = \varphi(t) \vd, \qquad
  \vp(t)= \frac{\dot\varphi \vd}{\sqrt{1-{\dot\varphi^2}d^2}},
\end{equation}
where we denoted $d^2=\vd^2$. Then, substitution to second half of  Hamilton's equations
\eqref{eq:eqham}, gives the second order differential equation
\begin{equation}
  \ddot\varphi = -\gamma\left(1-{\dot\varphi^2}d^2\right)^{3/2}
  \varphi^{k-1}.
  \label{eq:navarphi}
\end{equation}
for scalar function $\varphi$ provided  \eqref{eq:darboux} holds.

Equation \eqref{eq:navarphi} has the energy integral
\begin{equation}
  h=\frac{1}{d^2\sqrt{1-{\dot\varphi^2}d^2}}+
  \frac{\gamma}{k}\varphi^k.
  \label{eq:h}
\end{equation}

The variational equations along solution~\eqref{eq:part} have the form
\begin{equation}
  \label{eq:var}
  \begin{split}
    \dot \vx = &
    \sqrt{1-{\dot\varphi^2}d^2}
    \left[\vy-{\dot\varphi^2}(\vd\cdot\vy)\vd\right], \\
    \dot \vy = &-\varphi(t)^{k-2}V''(\vd) \vx.
  \end{split}
\end{equation}
Here we used the fact that $V''(\vq)$ is a homogeneous function of degree
$(k-2)$, hence $V''\left(\varphi(t) \vd\right)=\varphi^{k-2}(t)V''(\vd)$.
Hessian $V''(\vd)$ of the potential $V$ calculated at a Darboux point $\vd$ is a
symmetric matrix. Thus, in a generic case, there exists a complex orthogonal
$n\times n$ matrix $A$ such that the canonical change of variables
\begin{equation*}
  \vx = A \veta, \qquad  \vy = A \vxi,
\end{equation*}
transforms $V''(\vd)$ to its diagonal form with eigenvalues
$(\widehat\lambda_1,\ldots, \widehat\lambda_n)$. A Darboux point $\vd$ is an
eigenvector of $V''(\vd)$ corresponding to the eigenvalue
$\widehat\lambda_n=\gamma(k-1)$.  It is  transformed into
$\vD=A^T\vd=[0,\ldots,0,d]^T$. In effect, the variational
equations~\eqref{eq:var} are transformed to the form
\begin{equation}
  \label{eq:wariatys}
  \begin{aligned}
    &\dot \eta_i = \sqrt{1-{\dot\varphi(t)^2}d^2}\,\xi_i, && \dot \xi_i =
    -\widehat\lambda_i \varphi(t)^{k-2}\eta_i,
    \\
 &   \dot \eta_n = \left(1-{\dot\varphi(t)^2}d^2\right)^{3/2}\xi_i, && \dot \xi_n =
    -\widehat\lambda_n \varphi(t)^{k-2}\eta_n,
\end{aligned}
\end{equation}
where $i=1, \ldots, n-1$.

The second order equations for $\eta_i$, are as follows
\begin{equation}
  \label{eq:etai}
  \ddot \eta_i  +a \dot \eta_i+b \eta_i=0.
\end{equation}
The coefficients $a$ and $b$ for $i=1, \ldots, n-1$, have the form
\begin{equation}
  \begin{split}
    &a=-{d^2\gamma}\varphi(t)^{k-1}\dot{\varphi}(t)\sqrt{1-{\dot\varphi(t)^2}d^2},\\ &b=\widehat\lambda_i\varphi(t)^{k-2}\sqrt{1-{\dot\varphi(t)^2}d^2},
  \end{split}
  \label{eq:unve}
\end{equation}
while for $i=n$ they are
\begin{equation}
  \begin{split}
    &a=-{3d^2\gamma}\varphi(t)^{k-1}\dot{\varphi}(t)\sqrt{1-{\dot\varphi(t)^2}d^2},\\ & b=\widehat\lambda_n\varphi(t)^{k-2}\left(1-{\dot\varphi(t)^2}d^2\right)^{3/2}.
  \end{split}
  \label{eq:unvetang}
\end{equation}
The last equation, for $\eta_n$, describes the variations along the particular solution and it does not give any obstructions for the integrability, so we  consider only the first $(n-1)$ equations called normal variational equations.

Next,  we make the Yoshida transformation of the independent variable
\begin{equation}
  \label{eq:yoshi}
  t\longrightarrow z := \frac{\gamma d^2}{k} \varphi(t)^k, \quad \text{with} \quad \gamma dk\neq 0,
\end{equation}
see \cite{Yoshida:87::}.
We need the known formulae
\begin{equation}
  \frac{\rmd\phantom{t}}{\rmd t} x = {\dot z}\, x', \quad
  \frac{\rmd^2\phantom{t}}{\rmd t^2} x = {\dot z}^2\, x'' + \ddot z\, x', \qquad
  '\equiv\frac{\rmd \phantom{z}}{\rmd z},
\end{equation}
where, in our case, $\dot z^2$ and $\ddot z$ take the form
\begin{equation}
  \begin{split}
    &\dot z^2=\frac{\gamma  k \varphi^{k-2} z \left[(z-s)^2-1\right]}{(z-s )^2},\\
   & \ddot z=\frac{\gamma  \varphi^{k-2}\left[z+(k-1) \left(s +(z-s
          )^3\right)\right]}{(z-s )^3}.
  \end{split}
\end{equation}
Here $s=d^2 e$, while $e$ is a value of the energy integral  $h$ defined by~\eqref{eq:h},
corresponding to the selected particular solution~\eqref{eq:part}.  After
transformation~\eqref{eq:yoshi} equations~\eqref{eq:etai} read
\begin{equation}
  \eta_i''+p(z)  \eta_i'+q(z)\eta_i=0,\qquad i=1,\ldots,n-1,
  \label{eq:niezred}
\end{equation}
where
\begin{equation}
  \begin{split}
    p(z)=\frac{k-1}{k z}+\frac{z-s
    }{(z-s )^2-1},\,\,
    q(z)= \frac{\lambda  (s -z)}{k
      z \left[(z-s)^2-1\right]},
  \end{split}
  \label{eq:pq}
\end{equation}
and $\lambda=\lambda_i = {\widehat\lambda}_i/\gamma$.
All of the above  equations have the same form, the only dependence on index $i$ is through $\lambda_i$. Hence, we can consider just one equation
\begin{equation}
  \eta''+p(z)  \eta'+q(z)\eta=0,
  \label{eq:niezred1}
\end{equation}
with coefficients $p(z)$ and $q(z)$ given in~\eqref{eq:pq}.
For further analysis it is convenient to transform this equation into its
reduced form. We do this by making the following change of the dependent variable
\begin{equation}
  \label{eq:tran}
  \eta = w \exp\left[ -\frac{1}{2} \int_{z_0}^z p(u)\, du \right].
\end{equation}
Next, we obtain
\begin{equation}
  \label{eq:normalg}
  w'' = r(z) w, \qquad r(z) =\frac{1}{4}p(z)^2 + \frac{1}{2}p'(z) -q(z),
\end{equation}
where the explicit form of coefficient $r(z)$ is
\begin{multline}
  \label{eq:rrg}
  r(z) = \frac{k^2-1}{4k^2 z^2} +
  \frac{3}{16(z-s+1)^2} + \frac{3}{16(z-s-1)^2} \\
+ \frac{4s(1-k+2\lambda)-z(4-5k+8\lambda)}{8kz[(z-s)^2-1]}.
\end{multline}
The above shows that for generic values of the parameters, the  reduced equation
has four regular singular points at $z=0$,   $z=s\pm 1$, and  at $z=\infty$.

Summing up, we show that the normal variational equations are a direct product of
$(n-1)$ second order Fuchsian equations of the form \eqref{eq:normalg}. For its solvability analysis, we use the  Kovacic
algorithm~\cite{Kovacic:86::}. At the same time, we can
answer the question whether the identity component of its differential Galois
group is Abelian, that is, if the necessary conditions for the integrability
given by Theorem~\ref{thm:mo} are fulfilled. The true difficulty
is connected with the fact that the system depends on three parameters: $k, \lambda$, and $s$. However,
we have the freedom to select a value  of  parameter $s$. In fact, as $s=d^2 e$, so
choosing the energy~$e$ of the particular solution, we can change the value of
$s$ arbitrarily.

\subsection{Generic energy level}
\label{ssec:genericlevel}

Let us first consider the reduced normal variational equation~\eqref{eq:normalg},
at a generic value of the
energy with~$e\neq\pm d^{-2}$. Since this is a linear second differential
equation with rational coefficients, we determine its differential Galois group with the help of the Kovacic algorithm~\cite{Kovacic:86::}.
This algorithm decides, whether second order linear differential equation~\eqref{eq:normalg}, is solvable in a class of Liouvillian functions, and,  as a
by-product,  it enables to determine the differential Galois group of~\eqref{eq:normalg}. The algorithm consists of
four cases depending on the form of solutions and respective differential Galois
groups. In the first three cases, the equation is solvable, and its solutions
depend on a certain polynomial. The degree  of this polynomial depends only  on
$k$. The application of this algorithm, in our case,  consists of checking that
the degree of the mentioned polynomial is a non-negative integer. This analysis gives the following proposition.

\begin{proposition}
  \label{prop:MoRa}
  For $s\neq 1$, if  the identity component of the differential Galois group of equation \eqref{eq:normalg} with coefficient given in \eqref{eq:rrg} is solvable, then  pair $(k,\lambda_i)$ for $i=1,\ldots,n-1$, belongs to an item
  of the following list
  \begin{equation*}
    \begin{array}{cl}
      \hline
        k & \lambda  \\[0.5em]
      \hline
     k & (1 + p) (1 + k p), \\[0.5em]
     k &\frac{1}{4}(3 + 2 p) (2 + k + 2 k p)\\[0.5em]
     k & \frac{2 k-1}{4 k}+\frac{k}{16} \left(p^2-4\right), \\[0.5em]
     1 & \frac{p^2}{16}, \quad \frac{p^2}{144}, \quad\frac{p^2}{100}, \quad\frac{p^2}{64},\\[0.9em]
     2 & \frac{1}{8} \left(p^2-1\right), \quad \frac{p^2}{72}-\frac{1}{8}, \quad
   \frac{p^2}{50}-\frac{1}{8},\quad\frac{p^2}{32}-\frac{1}{8},\\[0.9em]
   3 & \frac{3 p^2}{64}-\frac{1}{3}, \quad\frac{p^2}{48}-\frac{1}{3},\quad \frac{3p^2}{100}-\frac{1}{3}, \\[0.9em]
   4 & \frac{1}{16} \left(p^2-9\right), \quad \frac{p^2}{36}-\frac{9}{16},\quad \frac{p^2}{25}-\frac{9}{16}, \\[0.9em]
   5 &\frac{5 p^2}{144}-\frac{4}{5}, \quad \frac{p^2}{20}-\frac{4}{5}, \quad \frac{5 p^2}{64}-\frac{4}{5},  \\[0.9em]
   6& \frac{p^2}{24}-\frac{25}{24}, \quad \frac{3 p^2}{32}-\frac{25}{24}, \quad \frac{3 p^2}{50}-\frac{25}{24}, \\[0.9em]
    \end{array}
\end{equation*}
  \begin{equation*}
    \begin{array}{cl}
    -1 & \frac{1}{16} \left(16-p^2\right),\quad 1-\frac{p^2}{144},\quad 1-\frac{p^2}{100},\quad 1-\frac{p^2}{64}, \\[0.9em]
-2 & \frac{1}{8}\left(9-p^2\right),\quad \frac{9}{8}-\frac{p^2}{72},\quad \frac{9}{8}-\frac{p^2}{50} ,\quad\frac{9}{8}-\frac{p^2}{32},\\[0.9em]
 -3 & \frac{4}{3}-\frac{3 p^2}{64},\quad\frac{4}{3}-\frac{p^2}{48},\quad \frac{4}{3}-\frac{3p^2}{100},\\[0.9em]
-4 & \frac{1}{16} \left(25-p^2\right),\quad \frac{25}{16}-\frac{p^2}{36},\quad \frac{25}{16}-\frac{p^2}{25},\\[0.9em]
-5 & \frac{9}{5}-\frac{5 p^2}{144},\quad\frac{9}{5}-\frac{p^2}{20}, \quad \frac{9}{5}-\frac{5p^2}{64}, \\[0.9em]
 -6 & \frac{49}{24}-\frac{p^2}{24},\quad\frac{49}{24}-\frac{3 p^2}{32},\quad \frac{49}{24}-\frac{3 p^2}{50}.\\[0.9em]
\hline
    \end{array}
\end{equation*}
   where $p$ is an integer and $k\neq 0$.
\end{proposition}
 We immediately notice from the above list, that if the system is integrable, then
all eigenvalues  of the Hessian are rational numbers. Of course, we know it from
Theorem~\ref{thm:moraleshomo}, however now we know that these numbers are either
integers or have specific numerators which are different from those given in Theorem~\ref{thm:moraleshomo}.

\subsection{ Special energy level}
\label{ssec:special}
Choosing  the energy of the particular solution as $e=d^{-2}$,   we  got
$s=1$, and so, two singular points at $z=0$ and at $z=s-1$ merge.   In effect,
we got a system with only three regular singularities $z=0$, $z=2$, $z=\infty$.
So, equation~\eqref{eq:normalg}  is the Riemann $P$-equation, \cite[Ch.X]{Whittaker:35::}. Conventionally  its
singular points are located at $z=0$, $z=1$, and in $z=\infty$. We achieve this
by making the change of the variable $z \mapsto 2 z$ in equation~\eqref{eq:normalg}. After this shift,  we obtain
\begin{equation}
  \label{eq:ourRiemann}
  w''= r(z) w,
\end{equation}
where now
\begin{equation}
  \label{eq:rrRiemann}
  r(z)= \frac{1}{4}\left(\dfrac{\rho^2-1}{z^2}+\dfrac{\sigma^2-1}{(z-1)^2}+
  \dfrac{1 +\tau^2 -\rho^2-\sigma^2}{z(z-1)}\right),
\end{equation}
and $\rho$, $\sigma$ and $\tau$ are the differences of exponents at singular
points at $z=0$, $z=1$ and in $z=\infty$, respectively. In our case, they are as follows
\begin{equation}
  \label{eq:rst}
  \begin{split}
    \rho=&\frac{\sqrt{(k-2)^2+8 k \lambda}}{2 k},\\ \sigma=&\frac{1}{2},\\
    \tau=&\frac{\sqrt{(k-1)^2+4 k \lambda }}{k}.
  \end{split}
\end{equation}
This simplification profits. Necessary and sufficient conditions guarantying
the solvability of the Riemann $P$-equation are known and they are given by the
Kimura theorem \cite{Kimura:69::}, see Appendix. For a more detailed analysis,
see~\cite{mp:20::f}. Using this theorem, we  deduce  that  our
equation~\eqref{eq:ourRiemann}, has the following property.
\begin{proposition}
If the identity component of the differential Galois group of equation
\eqref{eq:ourRiemann} is solvable, then   eigenvalues $\lambda=\lambda_i$  with
$i=1, \ldots n-1$,  are given by
\begin{enumerate}
\item $\lambda = f_\pm(k,p) $, or
\item $\lambda=\tfrac{1}{2}\left[\tfrac{k-1}{ k}+ k p (p+1)\right]$, or
\item $\lambda=\tfrac{2 k-1}{4 k}+\tfrac{1}{16} k (4 p (p+1)-3)$,
\end{enumerate}
where $p$ is an arbitrary integer and $k\neq 0$.
\label{prop:zhyper}
\end{proposition}
The analysis which leads to the above statements  is straightforward but quite
long, this is why we do not present it here.

Let us notice that the above proposition and
Theorem~\ref{thm:moraleshomo} are  both deduced from the Kimura Theorem \ref{kimura}, see Appendix.
However, for the non-relativistic case Theorem~\ref{thm:moraleshomo} specifies
17 cases,  while for the relativistic version of the system the above
proposition distinguishes only $4$ cases.

\subsection{Final steps}
\label{sec:final}

Let us assume that our Hamiltonian system \eqref{eq:eqham} is integrable in the
Liouville sense.  Then, by  Theorem~\ref{thm:mo} for an arbitrary particular
solution, the identity component of the differential Galois group of the
variational equations is Abelian. Our reasoning is based on the following
implication: if the identity component of the differential Galois group of
\eqref{eq:var} is Abelian, then the identity component of the differential
Galois group of \eqref{eq:niezred} and also its reduced form \eqref{eq:normalg}
is also Abelian.
Moreover, the choice of the energy level of the particular solution gives two
different sets of necessary conditions formulated in Proposition
\ref{prop:zhyper} and in Proposition \ref{prop:MoRa}. They  should be satisfied concomitantly.

At first, notice that, if the system is integrable, then by
Proposition~\ref{prop:MoRa}, all non-trivial eigenvalues $\lambda_i$  are rational.
Thus, the numbers $f_\pm(k,p)$ in Proposition~\ref{prop:zhyper} have to be
rational. From the definition of these numbers~\eqref{eq:fpm}, we deduce that
$m=\sqrt{4 k^2 p (2 p+1)+1}$ has to be a rational number. As both $k$ and $p$ are integers, $m^2=4 k^2 p (2 p+1)+1$ is an odd integer, therefore $m$ is odd.
Now, we have
\[
  f_\pm(k,p)= 3 k p (2 p+1)+
  \frac{1}{2}\left[1\pm (4 p+1) m\right].
\]
The expression in the square bracket is even number, so $f_\pm(k,p)$ are
integers. Hence, they are just elements of sets $\scJ_\pm$ defined
in~\eqref{eq:Jpm}.

The set of numbers defined with odd integer $p$ in the third line  of the table
Proposition~\ref{prop:MoRa}, coincides with the set given  in item 3 of
Proposition~\ref{prop:zhyper}.

Next, let us examine   the family  specified in item 2  of
Proposition~\ref{prop:zhyper} which are $\tfrac{k^2 p (1 + p)+k-1}{2k}$,
$p\in\Z$. Thus,  they are irreducible rational numbers of the form $\tfrac{s}{k}$
or $\tfrac{2s+1}{2k}$,  for  odd or even $k$, respectively. A lengthy and
laborious analysis shows that for integer $k$, $|k|>1$ these numbers do not
appear in appropriate families of Proposition~\ref{prop:MoRa}.

If $k=\pm1$, then elements of family of item 2  of
Proposition~\ref{prop:zhyper} are integers of the form
\begin{equation}
\lambda=\frac{1}{2}p(p+1),\mtext{or} \lambda=1-\frac{1}{2}p(p+1),\quad p\in\Z.
\label{eq:integi}
\end{equation}

For $k=1$ integer numbers   appear in  lines 1, 3 or 4, in the table given in
Proposition~\ref{prop:MoRa},   and they are perfect squares, that is
$\lambda=q^2$ for a certain integer $q$. Similarly, for $k=-1$  integer numbers
appear in  lines 1, 3 or 4, or 10 in  the table, and they are of the form
$\lambda=1-q^2$. Thus, for $k=1$  and $\lambda= q^2$ we have at the same time
$\lambda=\tfrac{1}{2}p(p+1)$, so $\lambda$ is square triangular number,   this
why $\lambda\in\scJ_{1}$. Similarly, for $k=-1$ we have that   $\lambda \in
\scJ_{-1}$.

This part of the reasoning is summarized in the following.
\begin{proposition}
\label{prop:gen+part}
  If the identity component of the differential Galois group of equation
  \eqref{eq:ourRiemann} is solvable, then   eigenvalues $\lambda=\lambda_i$  with
  $i=1, \ldots n-1$,  are given by
  \begin{enumerate}
  \item $\lambda \in \scJ_- \cup \scJ_+$, or
\item $\lambda=\tfrac{2 k-1}{4 k}+\tfrac{1}{16} k (4 p (p+1)-3)$, for $p\in \Z$, or
  \item for $\abs{k}=1$, $\lambda\in\scJ_k$.
  \end{enumerate}
\end{proposition}
In the last step, we recall Proposition~\ref{prop:relatnonrelat}, which says that
the integrability of a relativistic Hamiltonian system with a homoegeneous potential implies
the integrability of the corresponding non-relativistic system with the same
potential. Thus, now we have to find the intersection of  integrability conditions
formulated  in Proposition~\ref{prop:gen+part} with those  conditions for
the corresponding non-relativistic model  given in
Theorem~\ref{thm:moraleshomo}. In this way, we eliminate completely the
second item in the above proposition except for the case $\abs{k}\neq 2$ when Theorem~\ref{thm:moraleshomo} does not give any restriction.

The numbers given in the second case of Proposition~\ref{prop:gen+part}
can be written  as
\begin{equation}
\lambda=\frac{(2 k p-k+2) (2 k p+3
   k-2)}{16 k}.
   \label{eq:eleJ3}
\end{equation}
Thus, there are irreducible numbers of the form $\tfrac{2q+1}{16k}$ for $k$ odd. For $k$ even, we have to possibilities. 
If  $k=2 (2s+1)$ for $s\in\Z$, then $\lambda$ is irreducible rational number of
the form $\lambda = q/(2s+1)=q/(k/2)$ for a certain integer $q$; if $k=4 s$,
then $\lambda$ is irreducible rational number of the form $\lambda = (2q+1)/16s=
(2q+1)/(4k)$ for a certain integer $q$. Summarizing, the irreducible form of
$\lambda$ is
\begin{equation}
  \label{eq:cas}
  \lambda = \begin{cases}
    \frac{q}{16k} &  \mtext{for} k=2s+1, \\
    \frac{q}{4k} &  \mtext{for} k=4 s, \\
    \frac{q}{k/2} & \mtext{for}  k=2 (2s+1),
  \end{cases}
\end{equation}
where $q$ and $s$ are integers.

Now, we have to check if the numbers of these forms are listed in the table of
Theorem~\ref{thm:moraleshomo}. At first, we assume that $\abs{k}>2$. Rational, non-integers numbers are given in rows
3--9 in the table (in the third row for $\abs{k}>2$) of
Theorem~\ref{thm:moraleshomo}. The irreducible form of $\lambda$ in  the third
row of the table in this theorem  is either $ \lambda=(2q+1)/(2k)$ for even $k$,
or $ \lambda=q/k$ for odd $k$. Thus, they cannot coincide with numbers of the
form~\eqref{eq:cas}.

If $k=\pm 1$, then  admissible eigenvalues from Theorem~\ref{thm:moraleshomo} belong to
families in lines 2 and 3, are integer numbers and they have forms~\eqref{eq:integi}. Thus, they are
elements of sets $\scJ_{\pm1}$,  as  stated in the
Proposition~\ref{prop:gen+part} and in Theorem~\ref{thm:main}.

Finally, for $k=\pm2$  Theorem~\ref{thm:moraleshomo} does not give any
restriction. However,  as it is easy to verify, the second case in
Proposition~\ref{prop:gen+part} gives numbers of the forms \eqref{eq:integi}, and
thus they are elements of sets $\scJ_{\pm2}$ as it is claimed in
Theorem~\ref{thm:main}.

\section{Applications}
\label{sec:applications}

When considering the integrability of natural Hamiltonian  systems of the form
\eqref{eq:ham}, or \eqref{eq:hamnon-rel}, it is convenient   to  identify
potentials $V(\vq)$ and $V_A(\vq):=V(A\vq)$, where
$A\in\mathrm{PO}(2,\C)\subset\mathrm{GL}(2,\C)$ for all $A\in\mathrm{PO}(2,\C)$.
Here by $\mathrm{PO}(2,\C)$ we denote the two-dimensional complex projective
orthogonal group, that is the group $2\times 2$ complex matrices $A$ such that
$AA^T= \alpha E$  for a certain non-zero  $\alpha\in\C$. Clearly, integrability
of a particular potential from a given class implies the integrability of potentials
from this class.

Let us consider the relativistic systems  with two degrees of freedom and
homogeneous  potentials. Among  them  two families  are integrable  in both
relativistic and non-relativistic regimes, namely:
\begin{itemize}
\item  if $V=V(q_1)$ then  additional first integral is $F=p_2$;
\item if $V=V(r)$ with $r=\sqrt{q_1^2 +q_2^2}$, then
the  additional first integral is
$F=q_1p_2-q_2p_1$
\end{itemize}
Now, let us consider only polynomial homogeneous potentials. The case of polynomial
potentials of degree one is covered by the first of the above cases. Thus, we
start  with homogeneous potentials of degree $k=2$. In the non-relativistic case,
such potentials are always integrable. Let us consider such potentials in the
relativistic model.    A real homogeneous potential of degree 2 can be transformed
to a form of the anisotropic harmonic oscillator, which is given by
\begin{equation}
 \label{eq:os}
 V=q_1^2+\alpha q_2^2.
\end{equation}
 This potential has two Darboux points. The Hessians at these points have  non-trivial eigenvalues  $\alpha$ and $\alpha^{-1}$, respectively. We proved that if the system is
integrable, then both these numbers have to be integers, so $\alpha=\pm 1$.
Moreover,  $\alpha$ and $\alpha^{-1}$ must belong to
$\scJ_+\cup\scJ_-\cup\scJ_2$, see Theorem~\ref{thm:main}. As  $1\in
\scJ_+$, the case with $\alpha=1$ satisfies the necessary conditions for the integrability and is integrable. On the other hand,  $-1\notin
\scJ_+\cup\scJ_-\cup\scJ_2$, so for  $\alpha=-1$ the relativistic
system with potential~\eqref{eq:os} is not integrable.

As the next example, we consider cases of polynomial potentials of degree $k>2$.
The first elements of the set $\scJ_+\cup\scJ_-$, for respective $k=3,4,5,6$ are as follows
\begin{equation}
  \begin{split}
   &\{ 0, 1, 5, 40, 176, 1365,5985,\ldots\}, \\
   &\{0, 1, 10, 45, 351, 1540, 11935,\ldots\},\\
   &\{ 0, 1, 540, 1729, 18361, 58752,\ldots \},\\
   &\{0, 1, 21, 56, 736, 1925, 25025,\ldots\}.
  \end{split}
  \label{eq:zbiorkikeq3do6}
\end{equation}
In \cite{mp:04::d}, see also \cite{Hietarinta:83::}, it was proved that there are  six families of
integrable homogeneous  polynomial potentials of degree~$3$ with one or three  different Darboux points with the following non-trivial eigenvalues  of the Hessian
\begin{enumerate}
\item $V_1=q_1^3+\alpha q_2^3$, where $ \alpha\in \C^{\star}$ with $0,0,2$,
\item $V_2= \frac{1}{2}q_1^2 q_2 + q_2^3$  with $1/3,5,5$,
    \item  $V_3= \frac{1}{2}q_1^2 q_2 + \frac{8}{3}q_2^3,$ with $1/8,15,15$,
\item   $V_4 = \pm \frac{ \rmi\sqrt{3}}{18}q_1^3 + \frac{1}{2}q_1^2 q_2 + q_2^3,$ with $1/3,10/3,15$,
\item   $V_5=q_1^3,$  with $0$,
\item   $V_6=\frac{1}{3}(q_2\mp\rmi q_1)^{2}[q_2\pm2\rmi q_1]$ with  2,
\end{enumerate}
respectively. According to our main Theorem~\ref{thm:main} all non-trivial eigenvalues of  Hessians evaluated at all Darboux points must be  integer numbers belonging to the set $\scJ_+\cup\scJ_-$ given explicitly in the first line of equation \eqref{eq:zbiorkikeq3do6}. But this holds only for potential $V_5=q_1^3$ with only one Darboux point with nontrivial eigenvalue equals 0 and it is really integrable  in the relativistic model with additional first integral $F=p_2$. For remaining potentials at least one eigenvalue does not belong to the set $\scJ_+\cup\scJ_-$.

Similarly, a list of known integrable families of homogeneous  polynomial
potentials of degree $4$ and $5$ can be found in \cite{Hietarinta:83::}. In
\cite{mp:05::c} it was proved  that this list is complete. Using our
Theorem~\ref{thm:main} one can easily check that among them only potentials
$V=q_1^4$,  $V=(q_1^2+q_2^2)^2$  and $V=q_1^5$ are integrable in the
relativistic model, which is obvious.

For arbitrary integer $k\geq 2$ there exist two integrable non-relativistic
homogeneous polynomial potentials which are separable in parabolic and Cartesian coordinates, respectively. At first let us consider  potentials
separable in parabolic coordinates, which have the form
\begin{equation}
  V_\mathrm{p}=\sum_{i=0}^{[k/2]}2^{-2i}\binom{k-i}{i}q_1^{2i}q_2^{k-2i}.
\end{equation}
Since  they are separable, they are also integrable in the non-relativistic
model. However, in the relativistic regime all these potentials are not
integrable. To show this we notice that these potentials have $k$ Darboux
points, and the non-trivial eigenvalues of the Hessians at these points are
\[
  \frac{k-1}{2k},\underbrace{k+2,\ldots,k+2}_{k-1\rm\ times}.
\]
As for $k\geq 2$, number    $\tfrac{k-1}{2k}$ is not an integer, so the
necessary conditions for the integrability given in our Theorem~\ref{thm:main}
are not fulfilled, so the system is not integrable.

As a last but not least important example, let us consider  potentials separable in Cartesian coordinates, which have
the form
\begin{equation}
  \label{eq:sep_c}
  V_\mathrm{C} = q_1^k +\alpha q_2^k, \qquad \alpha\neq 0.
\end{equation}
These potentials have $k$ Darboux points, and the non-trivial eigenvalues of the Hessian's at these points are
\begin{equation}
0,0,\underbrace{k-1,\ldots,k-1}_{k-2\rm\ times}.
 \label{eq:sep_c_spectrum}
\end{equation}
Now, according to our theorem, if the system is integrable, then: $0 \in \scJ_+\cup\scJ_-$ and $k-1 \in \scJ_+\cup\scJ_-$. One can notice that
$0\in\scJ_-$ for any $k$ (just when we substitute $p=0$). If $k-1 \in \scJ_+\cup\scJ_-$, then from definition of sets $\scJ_+$
and $\scJ_-$, we deduce that equation
\begin{equation*}
  3 k p (2 p+1)+\tfrac{1}{2}\left[1\pm (4 p+1) \sqrt{4 k^2 p (2
    p+1)+1}\right]=k-1,
\end{equation*}
for given integer $k>2$ has to possess an integer solution $p\in\Z$.
From this equality we get
\[
  \left[1-k+p+kp(p-1) \right]\left[2-k-4p+4kp(p+2) \right]=0.
\]
Solution with respect $k$ gives
\[
  k=\frac{p+1}{1+p(1-p)}\quad\text{or}\quad k=\frac{2(2p-1)}{4 p (2 + p)-1}.
\]
It has integer solutions $(k,p)$, with integer $k\geq 3$, and $p\in \Z$ only for
$k=10$ that corresponds to $p=-2$ in the second solution. Summarizing, we proved
that, except $k=10$,  potentials~\eqref{eq:sep_c} are not integrable in the
relativistic model.
\begin{figure}[t]
  \begin{center}
    \includegraphics[width=0.35\textwidth]{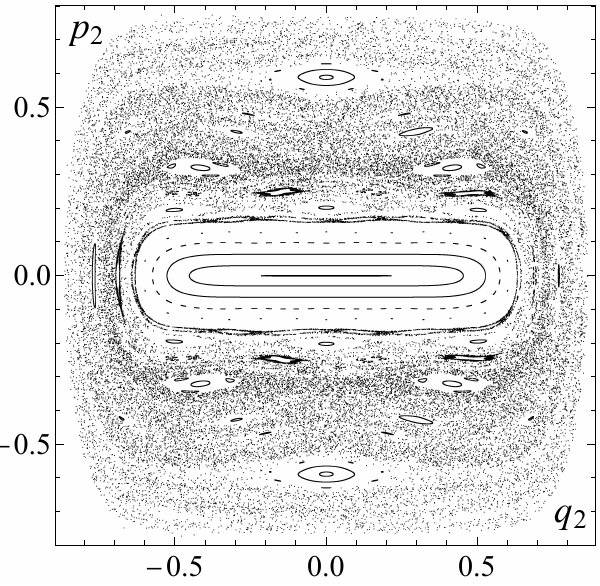}
    \caption{Poincar\'e section of the relativistic system with potential~\eqref{eq:sep_c} made for $k=10$, $\alpha=1$ at constant energy level $E=1.3$. The cross-section plane was specified as
      $q_1=0$, with $ p_1>0$. The plot shows the chaotic behaviour of the system, which confirms  its non-integrability.    \label{poin_last}}
  \end{center}
\end{figure}
Our Theorem~\ref{thm:main} gives necessary conditions for the
integrability, so  potentials~\eqref{eq:sep_c} with $k=10$ can be
non-integrable. Indeed, the Poincar\'e section, visible in Fig.~\ref{poin_last},
suggests the non-integrability of the system. We can prove this. Notice that in
Proposition~\ref{prop:MoRa} we extract  only a part of conditions guarantying
non-integrability. Knowing the explicit form of the potential one can perform
the Kovacic algorithm till the end i.e. not only to check that a polynomial that is included in a solution of the normal variational equation has a certain non-negative integer degree but to find it. Simple calculations, which we do not present here, show  that for $\alpha\neq 0$ such a polynomial does not exist. It implies that the identity component of the differential Galois group of  normal variational equations is not solvable, thus, in particular, is non-Abelian and hence the relativistic Hamiltonian system with potential~\eqref{eq:sep_c} for $k=10$ is not integrable.

\section{Remarks and comments}
\label{sec:recom}

Our main result, formulated in Theorem~\ref{thm:main} and its applications
presented in Section~\ref{sec:applications}, show that only in very exceptional cases relativistic versions of classical systems are integrable. It seems premature to conjecture that only systems with radial potentials and potentials
depending on one variable are integrable. Still, there are many open questions concerning the integrability in this context.

Let us remark here about an amazing fact concerning the necessary conditions for integrability given in Theorem~\ref{thm:main}. They state that  eigenvalues of the Hessian of the potential evaluated at a Darboux point have to be integer numbers  of
 very special forms.  It appears that these numbers, which belong to $\scJ_+\cup\scJ_-$,  can be expressed with the help of  solutions of
the famous Pell equation
\begin{equation}
  \label{eq:Pell}
  U^2 -D V^2 =1,
\end{equation}
see e.g.~\cite[Chap. 4]{Andreescu:15::}. In  our case the parameter $D$ takes the form $D=32k^2$. We proved that if numbers
$\lambda_n\in\scJ_{\pm}$, then they are given by the following recurrent relation
\begin{equation}
  \label{eq:kur}
  \lambda_{n+3}=a(\lambda_{n+2}-\lambda_{n+1})+\lambda_n,\qquad a=4U_1^2-1,
\end{equation}
where $(U_1,V_1)$ is the  fundamental solution of equation~\eqref{eq:Pell}.
For these numbers, one can find explicit formulae
\begin{equation}
  \lambda_n = \frac{-128 k^2(k^2-8k +6)+\eta_{\mp} Z_- +\eta_{\pm} Z_+}{2048 k^3},
\end{equation}
where
\begin{equation}
  Z_\pm = \left(X_0\pm 4 \sqrt{2} k Y_0\right)^2 \left(U_1\pm 4 \sqrt{2} k V_1\right)^{2 n}.
\end{equation}
Here $(X_0,Y_0)$ is a particular  solution of the general Pell equation
\begin{equation}
  X^2 - D Y^2 = N, \qquad  N= 64k^2(k^2-2),
\end{equation}
and $\eta_{\pm}= 3\pm 2 \sqrt{2}$.
The above formulae are important for theoretical investigations.

Here we mention
one open problem. In \cite{mp:05::c} it was shown that a generic polynomial potential of
degree $k$  has $k$ Darboux points $\vd_i$, and  the non-trivial eigenvalues of
its  scaled Hessian $\lambda_i= \gamma^{-1}\tr V''(\vd_i)-(k-1)$, satisfy the following universal relation
\begin{equation}
  \sum_{i=1}^k \frac{1}{\lambda_i -1}=-1,
  \label{eq:relka3}
\end{equation}
and this property is still valid in a relativistic regime. If the potential is integrable, then we know that all eigenvalues are integers.
Thus, the question is what are integer solutions of equation \eqref{eq:relka3} with $\lambda_i\in \scJ_+\cup\scJ_-$ for $i=1, \ldots, k$ for $k\geq3$.  As a
matter of fact, we found just one such solution  (for $k=10$) with nontrivial eigenvalues given in~\eqref{eq:sep_c_spectrum}. Nevertheless, we have shown that relativistic Hamilton equations with the corresponding potential \eqref{eq:sep_c}  are not integrable.  We conjecture that there are no other solutions, so  in the relativistic model, all generic polynomial potentials are not integrable. For $n\geq 3$  degrees of freedom a generic polynomial potential has $N=\tfrac{(k-1)^n-1}{k-2}$ Darboux points $\vd_i$ with $n-1$ non-trivial eigenvalues $\lambda_1(\vd_i),\ldots,\lambda_{n-1}(\vd_i)$ of the scaled Hessian $V''(\vd_i)$. Then between the non-trivial eigenvalues  at these points exist  $n$ universal relations, see \cite{mp:09::a}. Among them, one
\begin{equation}
    \begin{split}
        &\sum_{i=1}^N\left(\frac{1}{\lambda_1(\vd_i)-1}+\cdots+\frac{1}{\lambda_{n-1}(\vd_i)-1}\right)\\ &=-\dfrac{(k-1)^n-n(k-2)-1}{(k-2)^2},
    \end{split}
\end{equation}
is the generalisation of this in \eqref{eq:relka3},  and using it one can expect similar results as for $n=2$.

\section{Discussion and conclusions}
\label{sec:disc}
The article is devoted to the integrability analysis of Hamilton equations \eqref{eq:eqham} generated by Hamiltonian \eqref{eq:ham} describing a
relativistic particle moving in an external potential $V(\vq)$ in the limit of a
weak external field. Because the kinetic energy is no longer a quadratic form in
the momenta these equations differ significantly from their non-relativistic
counterparts.

We restrict our analysis to relativistic Hamiltonian systems with homogeneous potentials of an integer non-zero degree $k$. We noticed a direct relation between the integrability of relativistic and corresponding non-relativistic Hamiltonian
systems formulated in Proposition~\ref{prop:relatnonrelat}  and
Theorem~\ref{thm:main}. Proposition~\ref{prop:relatnonrelat} is based on the
expansion \eqref{eq:serh} of  relativistic Hamiltonian into a power
series of terms that are weigh-homogeneous functions and the lowest term of
weight-degree $2k$ is exactly the corresponding non-relativistic Hamiltonian
with this potential. This implies that for the integrability of the relativistic
system the integrability of the corresponding non-relativistic one is necessary.
Expansion~\eqref{eq:serh} also explains the observation that a relativistic
Hamiltonian system with a potential that is integrable in non-relativistic framework is  usually  non-integrable. This is the case for example separable
potentials. Namely, relativistic Hamiltonian systems can be considered as
perturbations of non-relativistic ones.  As it is well-known from the KAM theory
perturbations of integrable systems usually destroy integrability. However, in a
classical setting, by a perturbation we understand a small change  of the
potential.  However, in the relativistic version of perturbation consists of
a complete change of the kinetic energy term in the Hamiltonian function.  Clearly,
the consequences of this change need a deeper investigation. Let us mention here
simple observations. First of all, the relativistic version~\eqref{eq:ser} of
the classical system~\eqref{eq:hamnon-rel}  has the same equilibria. Moreover,
in both cases they are of the same stability type, even more, linearizations at
these points coincide.  It is no longer true with periodic solutions. Examples
given  in Section II show this explicitly.

The main result of this paper is Theorem~\ref{thm:main}, which contains
necessary integrability conditions formulated in admissible values of
non-trivial eigenvalues of rescaled Hessian matrix $\gamma^{-1}V''(\vd)$.
Namely, all of them must be integer numbers of appropriate form depending on
$k$. Let us notice that these conditions are much stronger than those for
non-relativistic systems with homogeneous potentials given in
Theorem~\ref{thm:moraleshomo}, where non-integer rational eigenvalues of
$\gamma^{-1}V''(\vd)$ were also admissible. The strength of obtained conditions
stems from the fact that they were derived by the intersection of conditions obtained
from analysis of the differential Galois group of variational equations along
two different particular solutions and also conditions for non-relativistic
potentials.

Application of the obtained conditions to homogeneous polynomial potentials of small degree  allows us to presume that the only
relativistic Hamiltonian systems with such potentials are radial potentials
$V(r)$ and those which depend only one coordinate $V(q_1)$.

Obviously in applications the problem of the integrability of relativistic systems
\eqref{eq:eqham} with  non-homogeneous potentials appears. The differential
Galois obstructions formulated in Theorem~\ref{thm:mo} can be applied provided a particular, non-equilibrium solution is known. In the case when a non-homogeneous potential is a sum of a radial potential
$r^{k}=(q_1^2+\cdots+q_n^2)^{k/2}$, $k\in\Z$ and a homogeneous potential $V_h(\vq)$ of
degree $l$, $l\neq k$,  i.e. $V(\vq)=r^{k}+V_h(\vq)$, that admits a Darboux
point $\vd$ satisfying \eqref{eq:darboux}, then still one can  construct a particular solution of Hamilton equations with  non-homogeneous potential
$V(\vq)$ by means of $\vd$. We do not analyse this class of potentials with
arbitrary  $r^{k}$ and $V_h(\vq)$ because variational equations are too
complicated but for particular radial and homogeneous potentials such analysis
is possible.

\appendix
\section{Riemann $P$-equation}
\label{app:riemann}

The Riemann $P$-equation \cite{Whittaker:35::}, is the most general
second-order differential equation with three regular singularities.
If we place these singularities at $z=0,1,\infty$, then it has the
form
\begin{multline}
\dfrac{\mathrm{d}^2\xi}{\mathrm{d}z^2}+\left(\dfrac{1-\alpha-\alpha'}{z}+
\dfrac{1-\gamma-\gamma'}{z-1}\right)\dfrac{\mathrm{d}\xi}{\mathrm{d}z}\\
+
\left(\dfrac{\alpha\alpha'}{z^2}+\dfrac{\gamma\gamma'}{(z-1)^2}+
\dfrac{\beta\beta'-\alpha\alpha'-\gamma\gamma'}{z(z-1)}\right)\xi=0,
\label{eq:riemann}
\end{multline}
where $(\alpha,\alpha')$, $(\gamma,\gamma')$ and $(\beta,\beta')$ are the
exponents at singular points. Exponents satisfy the Fuchs relation
\[
\alpha+\alpha'+\gamma+\gamma'+\beta+\beta'=1.
\]
We denote differences of exponents by
\[
\rho=\alpha-\alpha',\qquad\sigma=\gamma-\gamma',\qquad\tau=\beta-\beta'.
\]

For equation \eqref{eq:riemann} the necessary and sufficient conditions for  the solvability of the identity
component of its differential Galois group  are given in the following theorem due to Kimura \cite{Kimura:69::}, see
also \cite{Morales:99::c}.
\begin{theorem}[Kimura]
  The identity component of the differential Galois group of
  equation~\eqref{eq:riemann} is solvable if and only if
\begin{itemize}
\item[\emph{I:}] at least one of the four numbers $\rho+\tau+\sigma$,
  $-\rho+\tau+\sigma$, $\rho-\tau+\sigma$, $\rho+\tau-\sigma$ is an odd
  integer, or
\item[\emph{II:}] the numbers $\rho$ or $-\rho$ and $\tau$ or $-\tau$ and
  $\sigma$ or $-\sigma$ belong (in arbitrary order) to some of the
  following fifteen families forming the so-called Schwarz’s Table~\ref{tab:sch_app}.\\[0.5em]
\begin{table}[hb]
	\begin{center}
 \begin{ruledtabular}
 \begin{tabular}{l@{\hspace{0.5cm}}l@{\hspace{0.5cm}}l@{\hspace{0.5cm}}l@{\hspace{0.5cm}}l}
        \text{1}&$1/2+r$&$1/2+q$&arbitrary &\\
        \text{2}&$1/2+r$&$1/3+q$&$1/3+p$&\\
        \text{3}&$2/3+r$&$1/3+q$&$1/3+p$&$r+q+p$ even\\
        \text{4}&$1/2+r$&$1/3+q$&$1/4+p$&\\
        \text{5}&$2/3+r$&$1/4+q$&$1/4+p$&$r+q+p$ even\\
        \text{6}&$1/2+r$&$1/3+q$&$1/5+p$&\\
        \text{7}&$2/5+r$&$1/3+q$&$1/3+p$&$r+q+p$ even\\
        \text{8}&$2/3+r$&$1/5+q$&$1/5+p$&$r+q+p$ even\\
        \text{9}&$1/2+r$&$2/5+q$&$1/5+p$&\\
        \text{10}&$3/5+r$&$1/3+q$&$1/5+p$&$r+q+p$ even\\
        \text{11}&$2/5+r$&$2/5+q$&$2/5+p$&$r+q+p$ even\\
        \text{12}&$2/3+r$&$1/3+q$&$1/5+p$&$r+q+p$ even\\
        \text{13}&$4/5+r$&$1/5+q$&$1/5+q$&$r+q+p$ even\\
        \text{14}&$1/2+r$&$2/5+q$&$1/3+p$& \\
        \text{15}&$3/5+r$&$2/5+q$&$1/3+p$&$r+q+p$ even\\
  \end{tabular}
  \end{ruledtabular}
   \caption{Schwarz's table. Here $r,q,p\in\Z$. \label{tab:sch_app}}
\end{center}
\end{table}
\end{itemize}
\label{kimura}
\end{theorem}
\section*{Acknowledgements}
This research has been  founded by The
National Science Center of Poland under Grant No.
2020/39/D/ST1/01632. For the purpose of Open
Access, the authors have applied a CC-BY public
copyright license to any Author Accepted Manuscript
(AAM) version arising from this submission.

\section*{Data Availability Statement}
Data sharing is not applicable to this article as no new data were created or analyzed in this study.

\bibliographystyle{unsrtnat}
\def\cprime{$'$} \def\cydot{\leavevmode\raise.4ex\hbox{.}} \def\cprime{$'$}
  \def\polhk#1{\setbox0=\hbox{#1}{\ooalign{\hidewidth
  \lower1.5ex\hbox{`}\hidewidth\crcr\unhbox0}}} \newcommand{\noopsort}[1]{}

\end{document}